\pdfoutput=1

\documentclass[12pt,a4paper]{article}

\usepackage{ifthen} 
\newboolean{pdflatex}
\setboolean{pdflatex}{true} 

\newboolean{articletitles}
\setboolean{articletitles}{true} 

\newboolean{uprightparticles}
\setboolean{uprightparticles}{false} 

\def\papertitle{Towards an amplitude analysis of the decay \LbpKGam } 

\usepackage[top=1in, bottom=1.25in, left=1in, right=1in]{geometry}

\usepackage{microtype}
\usepackage{lineno}  
\usepackage{xspace} 
\usepackage{caption} 

\usepackage{graphicx}  
\usepackage{color}
\usepackage{colortbl}
\graphicspath{{./figs/}} 
\DeclareGraphicsExtensions{.pdf,.PDF,png,.PNG}

\usepackage{amsmath} 
\usepackage{amssymb}
\usepackage{amsfonts}
\usepackage{upgreek} 

\newcommand*\patchAmsMathEnvironmentForLineno[1]{%
\expandafter\let\csname old#1\expandafter\endcsname\csname #1\endcsname
\expandafter\let\csname oldend#1\expandafter\endcsname\csname
end#1\endcsname
 \renewenvironment{#1}%
   {\linenomath\csname old#1\endcsname}%
   {\csname oldend#1\endcsname\endlinenomath}%
}
\newcommand*\patchBothAmsMathEnvironmentsForLineno[1]{%
  \patchAmsMathEnvironmentForLineno{#1}%
  \patchAmsMathEnvironmentForLineno{#1*}%
}
\AtBeginDocument{%
\patchBothAmsMathEnvironmentsForLineno{equation}%
\patchBothAmsMathEnvironmentsForLineno{align}%
\patchBothAmsMathEnvironmentsForLineno{flalign}%
\patchBothAmsMathEnvironmentsForLineno{alignat}%
\patchBothAmsMathEnvironmentsForLineno{gather}%
\patchBothAmsMathEnvironmentsForLineno{multline}%
\patchBothAmsMathEnvironmentsForLineno{eqnarray}%
}

\usepackage{hyperxmp}

\usepackage[pdftex]{hyperref}

\usepackage[colorinlistoftodos,textsize=scriptsize]{todonotes}

\usepackage[all]{hypcap} 

\usepackage{xspace} 
\usepackage{upgreek}

\def\MagUp {\mbox{\em Mag\kern -0.05em Up}\xspace}

\ifthenelse{\boolean{uprightparticles}}%
{
 
 \def\Pgamma      {\ensuremath{\upgamma}\xspace}

 \def\Ppsi        {\ensuremath{\uppsi}\xspace}

 \def\PDelta      {\ensuremath{\Delta}\xspace}                 
 \def\PXi         {\ensuremath{\Xi}\xspace}                 
 \def\PLambda     {\ensuremath{\Lambda}\xspace}                 
 \def\PSigma      {\ensuremath{\Sigma}\xspace}                 
 \def\POmega      {\ensuremath{\Omega}\xspace}                 
 \def\PUpsilon    {\ensuremath{\Upsilon}\xspace}

 \def\PB      {\ensuremath{\mathrm{B}}\xspace}                 
                  
 \def\PD      {\ensuremath{\mathrm{D}}\xspace}

 \def\PJ      {\ensuremath{\mathrm{J}}\xspace}                 
 \def\PK      {\ensuremath{\mathrm{K}}\xspace}

 \def\Pb      {\ensuremath{\mathrm{b}}\xspace}

 \def\Pi      {\ensuremath{\mathrm{i}}\xspace}

 \def\Pp      {\ensuremath{\mathrm{p}}\xspace}

 \def\Ps      {\ensuremath{\mathrm{s}}\xspace}

 \def\thebaroffset{0.0em}
}
{
 
 \def\Pgamma      {\ensuremath{\gamma}\xspace}

 \def\Ppsi        {\ensuremath{\psi}\xspace}                 
                  
 \mathchardef\PDelta="7101
 \mathchardef\PXi="7104
 \mathchardef\PLambda="7103
 \mathchardef\PSigma="7106
 \mathchardef\POmega="710A
 \mathchardef\PUpsilon="7107
                  
 \def\PB      {\ensuremath{B}\xspace}                 
                  
 \def\PD      {\ensuremath{D}\xspace}

 \def\PJ      {\ensuremath{J}\xspace}                 
 \def\PK      {\ensuremath{K}\xspace}

 \def\Pb      {\ensuremath{b}\xspace}

 \def\Pi      {\ensuremath{i}\xspace}

 \def\Pp      {\ensuremath{p}\xspace}

 \def\Ps      {\ensuremath{s}\xspace}

 \def\thebaroffset{0.18em}
}
\newcommand{\offsetoverline}[2][\thebaroffset]{\kern #1\overline{\kern -#1 #2}}%

\makeatletter
\ifcase \@ptsize \relax
  \newcommand{\miniscule}{\@setfontsize\miniscule{4}{5}}
\or
  \newcommand{\miniscule}{\@setfontsize\miniscule{5}{6}}
\or
  \newcommand{\miniscule}{\@setfontsize\miniscule{5}{6}}
\fi
\makeatother

\DeclareRobustCommand{\optbar}[1]{\shortstack{{\miniscule (\rule[.5ex]{1.25em}{.18mm})}
  \\ [-.7ex] $#1$}}

\def\g      {{\ensuremath{\Pgamma}}\xspace}

\def\squark    {{\ensuremath{\Ps}}\xspace}

\def\bquark    {{\ensuremath{\Pb}}\xspace}

\def\kaon    {{\ensuremath{\PK}}\xspace}

\def\KorKbar {\kern \thebaroffset\optbar{\kern -\thebaroffset \PK}{}\xspace}

\def\Km      {{\ensuremath{\kaon^-}}\xspace}

\def\DorDbar {\kern \thebaroffset\optbar{\kern -\thebaroffset \PD}\xspace}

\def\B       {{\ensuremath{\PB}}\xspace}

\def\BorBbar {\kern \thebaroffset\optbar{\kern -\thebaroffset \PB}\xspace}

\def\Bd      {{\ensuremath{\B^0}}\xspace}

\def\BdorBdbar {\kern \thebaroffset\optbar{\kern -\thebaroffset \Bd}\xspace}

\def\Bs      {{\ensuremath{\B^0_\squark}}\xspace}

\def\BsorBsbar {\kern \thebaroffset\optbar{\kern -\thebaroffset \Bs}\xspace}

\def\jpsi     {{\ensuremath{{\PJ\mskip -3mu/\mskip -2mu\Ppsi\mskip 2mu}}}\xspace}

\def\Y#1S{\ensuremath{\PUpsilon{(#1S)}}\xspace}

\def\proton      {{\ensuremath{\Pp}}\xspace}

\def\Lz          {{\ensuremath{\PLambda}}\xspace}

\def\LorLbar     {\kern \thebaroffset\optbar{\kern -\thebaroffset \PLambda}\xspace}

\def\Sigmares    {{\ensuremath{\PSigma}}\xspace}
\def\Sigmaz      {{\ensuremath{\Sigmares{}^0}}\xspace}

\def\Lb           {{\ensuremath{\Lz^0_\bquark}}\xspace}

\newcommand{\decay}[2]{\ensuremath{#1\!\to #2}\xspace} 

\def\to                 {\ensuremath{\rightarrow}\xspace}

\def\qsq       {{\ensuremath{q^2}}\xspace}

\def\AT#1     {\ensuremath{A_{\mathrm{T}}^{#1}}\xspace}           

\def\C#1      {\ensuremath{\mathcal{C}_{#1}}\xspace}                       
\def\Cp#1     {\ensuremath{\mathcal{C}_{#1}^{'}}\xspace}                    
\def\Ceff#1   {\ensuremath{\mathcal{C}_{#1}^{\mathrm{(eff)}}}\xspace}        
\def\Cpeff#1  {\ensuremath{\mathcal{C}_{#1}^{'\mathrm{(eff)}}}\xspace}       
\def\Ope#1    {\ensuremath{\mathcal{O}_{#1}}\xspace}                       
\def\Opep#1   {\ensuremath{\mathcal{O}_{#1}^{'}}\xspace}                    

\newcommand{\bra}[1]{\ensuremath{\langle #1|}}             
\newcommand{\ket}[1]{\ensuremath{|#1\rangle}}              
\newcommand{\braket}[2]{\ensuremath{\langle #1|#2\rangle}} 

\newcommand{\aunit}[1]{\ensuremath{\text{\,#1}}}       

\newcommand{\tev}{\aunit{Te\kern -0.1em V}\xspace}
\newcommand{\gev}{\aunit{Ge\kern -0.1em V}\xspace}
\newcommand{\mev}{\aunit{Me\kern -0.1em V}\xspace}
\newcommand{\kev}{\aunit{ke\kern -0.1em V}\xspace}
\newcommand{\ev}{\aunit{e\kern -0.1em V}\xspace}
\newcommand{\mevc}{\ensuremath{\aunit{Me\kern -0.1em V\!/}c}\xspace}
\newcommand{\gevc}{\ensuremath{\aunit{Ge\kern -0.1em V\!/}c}\xspace}
\newcommand{\mevcc}{\ensuremath{\aunit{Me\kern -0.1em V\!/}c^2}\xspace}
\newcommand{\gevcc}{\ensuremath{\aunit{Ge\kern -0.1em V\!/}c^2}\xspace}
\newcommand{\gevgevcccc}{\ensuremath{\gev^2\!/c^4}\xspace} 

\def\gsim{{~\raise.15em\hbox{$>$}\kern-.85em
          \lower.35em\hbox{$\sim$}~}\xspace}
\def\lsim{{~\raise.15em\hbox{$<$}\kern-.85em
          \lower.35em\hbox{$\sim$}~}\xspace}

\def\tell1  {TELL1\xspace}
\def\ukl1   {UKL1\xspace}

\newcommand{\etc}{\mbox{\itshape etc.}\xspace}

\def\LbpKGam     {\decay{\Lb}{\proton \Km \g}}
\def\LbpKJpsi    {\decay{\Lb}{\proton \Km \jpsi}}
\def\LbpKll   {\decay{\Lb}{\proton \Km \ell^{+} \ell^{-}}}

\usepackage{cite} 
\usepackage{mciteplus}

\usepackage{longtable} 
\newcommand{\LSt}{\Lambda^*}
\usepackage{physics}
\begin{document}

\renewcommand{\thefootnote}{\fnsymbol{footnote}}


\begin{titlepage}

\vspace*{-1.5cm}

\noindent
 \\

\vspace*{4.0cm}

{\normalfont\bfseries\boldmath\huge
\begin{center}
  \papertitle
\end{center}
}

\vspace*{2.0cm}

\begin{center}
Johannes Albrecht$^1$, Yasmine Amhis$^2$, Anja Beck$^{1,2}$\footnote{Corresponding authors}, Carla Marin Benito$^{2\,*}$.
\bigskip\\
{\normalfont\itshape\footnotesize
$ ^1$ Fakult{\"a}t Physik, Technische Universit{\"a}t Dortmund, Dortmund, Germany \\
$ ^2$ Universit\'e Paris-Saclay, CNRS/IN2P3, IJCLab, 91405 Orsay, France 
}
\end{center}

\vspace{\fill}

\begin{abstract}
  \noindent
The helicity formalism applied to the  radiative decay \LbpKGam  is presented for the first time in this paper.  The aim is to provide the necessary formalism  to be able to resolve the resonant $\proton\Km$ structures at the photon pole by means of an amplitude analysis. Experimental effects, such as resolution, are also discussed.  
\end{abstract}

\vspace*{2.0cm}
\vspace{\fill}

\end{titlepage}

\pagestyle{empty}  


\newpage
\setcounter{page}{2}
\mbox{~}

\cleardoublepage


\renewcommand{\thefootnote}{\arabic{footnote}}
\setcounter{footnote}{0}



\pagestyle{plain} 
\setcounter{page}{1}
\pagenumbering{arabic}


%



\newpage
\section{Introduction}
\label{sec:Introduction}
Rare decays of \bquark hadrons,  such as  $b \to s \ell^{+} \ell^{-}$,    are Flavour-Changing Neutral-Currents (FCNC), which are forbidden at tree level in the Standard Model (SM) and are thus highly suppressed.  As such, they are very sensitive to potential new particles that can enter virtually through loop-level processes or allow new tree-level diagrams, affecting properties of the decays such as branching fractions and angular distributions. The measurement of these processes allows to probe higher scales  than those accessible via direct searches. 
Lepton Universality (LU) tests, which are complementary to these measurements and 
provide theoretically very precise observables, show a deviation with respect to the universal  SM prediction~\cite{LHCb-PAPER-2014-024,LHCb-PAPER-2017-013}. 

Thanks to the abundant production of \bquark baryons at LHC, the LHCb collaboration reported the first test of LU using \LbpKll decays\footnote{Charge conjugation is implied throughout the text.}~\cite{Aaij:2019bzx}, in the dilepton mass-squared range \mbox{$0.1 < \qsq < 6.0  \gevgevcccc$}  and the $\proton\Km$ mass range \mbox{$m_{\proton\Km} < 2600 \mevcc$}.
Direct interpretations of this result are difficult given that the resonant structure of the $\proton\Km$ final state is not resolved in this region of \qsq.  
The spectrum is well known at high \qsq following
a first amplitude analysis of \LbpKJpsi decays, which led to the discovery of states compatible with pentaquarks~\cite{Aaij:2015tga}.
In order to be able to characterise the $\proton\Km$ spectrum at the photon pole, the helicity formalism is employed in this work to derive the necessary building blocks for an amplitude analysis of the \LbpKGam decay.

This paper is organised as follows; the general three body helicity amplitude formalism is presented in Section~\ref{sec:helicity}, which is then applied to the specific decay \LbpKGam in Section~\ref{sec:specificLbpKG}, where no assumptions on the \Lb polarisation nor the photon polarisation are made. Experimental aspects to be considered in an amplitude fit to experimental data are discussed in Section~\ref{sec:fitstodata}.


\section{General three-body helicity amplitude}
\label{sec:helicity}
We describe a three-body decay as two consecutive two-body decays with an isobar model. This section presents the general three-body helicity amplitude. After an introduction to the helicity formalism, the inclusion of the dynamical description of the isobar model is discussed. Finally, the differential decay rate for a three-body decay including the option for various decay chains is given.
\subsection{Helicity formalism for a three-body decay}\label{sec:helFormThreeBody}
\begin{figure}
    \centering
    \includegraphics[width=\textwidth]{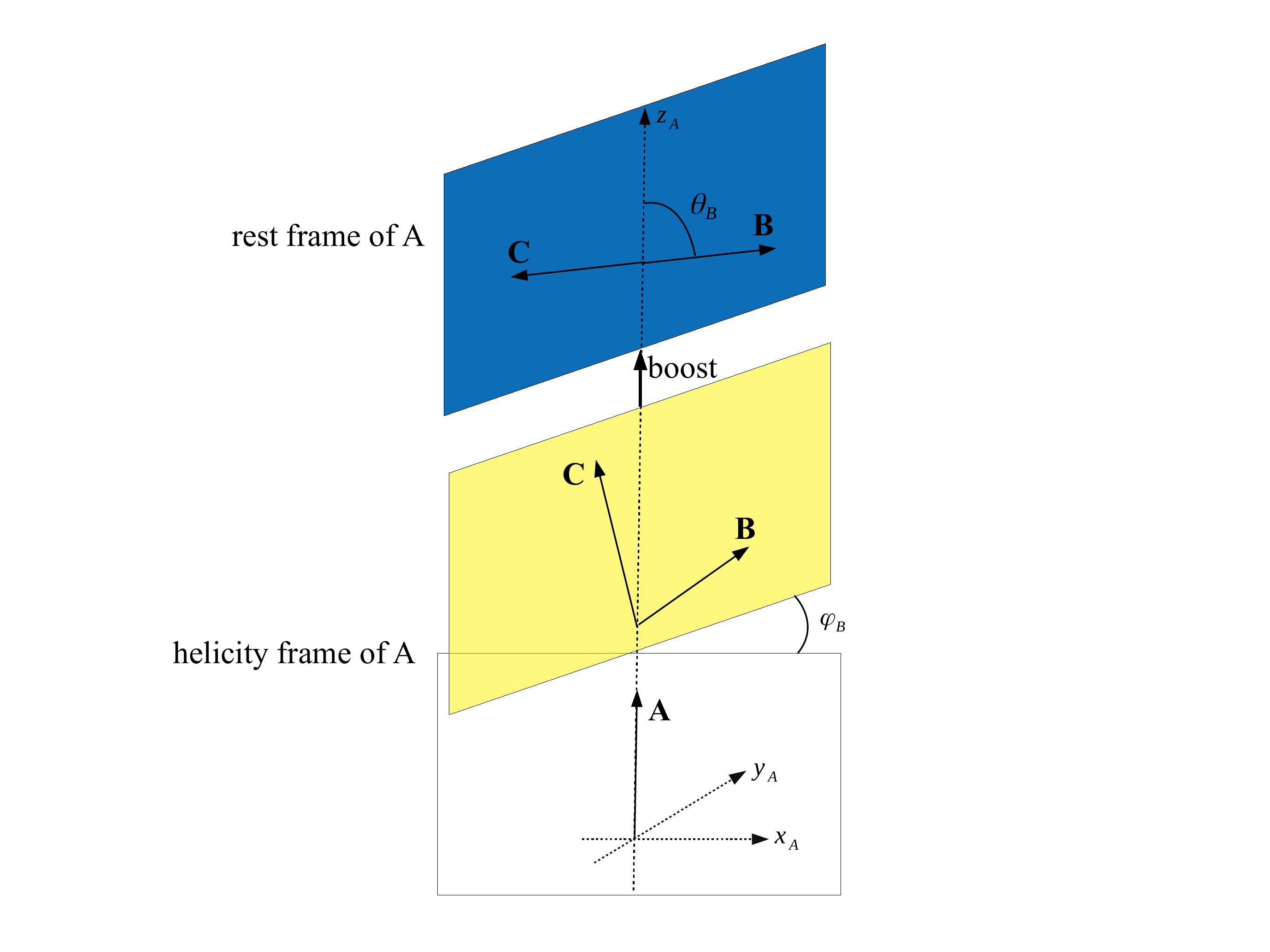}
    \caption{Visualisation of the decay angles in the decay $A\to BC$. The white/bottom plane is an arbitrary helicity frame of $A$, i.e.~the $z$ axis coincides with the momentum of $A$ and $x,y$ are arbitrary. The azimuthal angle of $B$, $\varphi_B$, describes how the decay plane (yellow/centre) lies within the arbitrary first frame. After a boost into the rest frame of $A$ (green/top plane), the polar angle of $B$, $\theta_B$, is defined.}
    \label{fig:rotationHelFrame}
\end{figure}

\begin{figure}
    \centering
    \includegraphics[width=\textwidth]{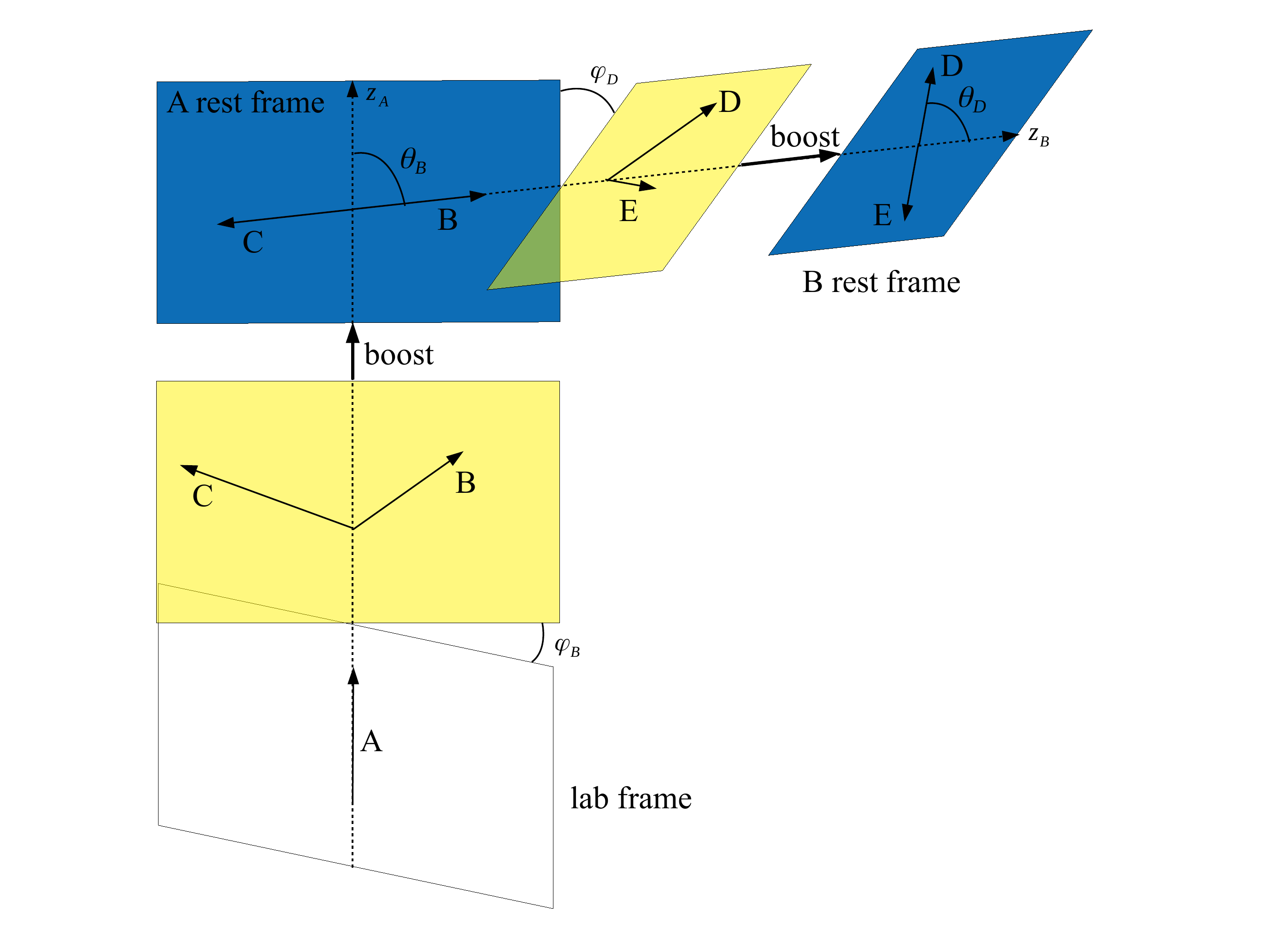}
    \caption{Definition of the decay angles for two consecutive two-body decays; the colour code follows that of Fig.~\ref{fig:rotationHelFrame}. The white/bottom plane, denoted ``lab frame", represents the arbitrary coordinate system in the helicity frame of $A$. The yellow planes (left centre and top centre) represent the decay planes in the parent helicity frame. The green planes (top left and top right) represent the parent rest frames and visualise the definition of the polar angles $\theta$.}
    \label{fig:rotationBothDecays}
\end{figure}

Using the helicity formalism, first derived in Ref.~\cite{JACOB1959404} and extensively discussed for example  in Refs.~\cite{Chung:186421,Richman:1984gh}, the decay of a particle $A$ with spin $J_A$ and helicity $\lambda_A$ to two child particles $B$ and $C$ ($A\to BC$) with helicities $\lambda_B$ and $\lambda_C$ is the product of a Wigner-D function $D^{J}_{mm'}(\alpha,\beta,\gamma)$ and a helicity coupling $H$:
\begin{equation*}
    D^{J_A}_{(\lambda_B-\lambda_C)\lambda_A}(\varphi_B,\theta_B,-\varphi_B)H^{A\to BC}_{\lambda_B,\lambda_C} \ .
\end{equation*}
The angles $\varphi_B$ and $\theta_B$ are the azimuthal and polar angle of child $B$, and thus describe the position of the decay axis in the parent coordinate system. In Fig.~\ref{fig:rotationHelFrame}, the rotation angles are illustrated; for a derivation and commentary on the choice of angles, see Appendix~\ref{app:helAmp}.
In order to model a consecutive decay $B\to DE$, this amplitude can easily be extended to
\begin{equation}\label{eq:helAmpThreeBody}
    \mathcal{A} =
	D^{J_A}_{(\lambda_B-\lambda_C)\lambda_A}(\varphi_B,\theta_B,-\varphi_B)H^{A\to BC}_{\lambda_B,\lambda_C}(-1)^{J_C-\lambda_C}D^{J_B}_{(\lambda_D-\lambda_E)\lambda_B}(\varphi_D,\theta_D,-\varphi_D)H^{B\to DE}_{\lambda_D,\lambda_E} (-1)^{J_E-\lambda_E} \ ,
\end{equation}
where the $(-1)^{J_X-\lambda_X}$ terms arise from the choice of the phase~\cite[Eq.~8]{Mikhasenko:2019rjf}.
For a visualisation of the angles used here, see Fig.~\ref{fig:rotationBothDecays}.
\subsection{Including dynamics}
The primary goal of the amplitude analysis of $\Lb\to(\LSt\to \proton\kaon^-)\gamma$ is to measure the contributions of the individual resonances such as $\Lz(1520)$, $\Lz(1690)$, \etc To this end, describing the resonance spectrum with an isobar model, where each resonance is represented by a relativistic Breit-Wigner function, is a common and convenient approach summarised in Ref.~\cite[Sec.~48.2]{PhysRevD.98.030001}. Using the particle labels introduced in the previous section, the standard parametrisation of this line shape for a resonance $B$ with mass $m_0$ and width $\Gamma_0$ is
\begin{align*}
    R(m_{DE}) &= \frac{1}{m_0^2-m_{DE}^2-im_0\Gamma(m_{DE})}, \\ \Gamma(m_{DE}) &= \Gamma_0\left( \frac{p}{p_0} \right)^{2l+1}\frac{m_0}{m_{DE}}\left[B_l(p,p_0)\right]^2 \ .
\end{align*}
Here, $p=p(m_{DE})$ is the absolute momentum of one of the children ($D$ or $E$) measured in the parent ($B$) rest frame, while $p_0=p(m_0)$ is the child's momentum at the resonance pole\footnote{The explicit formula is given in Eq.~\eqref{eq:momentum} in Appendix~\ref{app:helAmp}, where $m_{DE}$, respectively $m_0$, is the mass of the parent.}, $l$ is the orbital angular momentum between the two children $D$ and $E$, and $B_l$ is the Blatt-Weisskopf form factor \cite{Blatt:1952ije,PhysRevD.5.624}.

Additionally, the line shape $R(m_{DE})$ is accompanied by a Blatt-Weisskopf form factor and an angular momentum barrier for both of the decays to account for the suppression of high orbital angular momenta. The dynamical part of the decay rate is now
\begin{equation}\label{eq:dynamics}
    X(m_{DE}) = \left(\frac{q}{m_{A}}\right)^L\left(\frac{p}{m_0}\right)^lB_L(q,q_0)B_l(p,p_0)R(m_{DE}) \ ,
\end{equation}
with the orbital angular momentum $L$ between $B$ and $C$ and the absolute momentum $q$ of $B$ (or $C$) in the rest frame of $A$. In analogy to $p_0$, $q_0$ is the momentum at the resonance pole.

To incorporate the dynamics given by Eq.~\eqref{eq:dynamics} into the helicity amplitude in Eq.~\eqref{eq:helAmpThreeBody}, the couplings $H$ are transformed from the helicity basis to a basis defined by orbital angular momentum and total spin, the $LS$ basis. The technical aspects of this transformation and the inclusion of the dynamical part are discussed in  Appendix~\ref{app:dynamics}; for completeness, a comprehensive review on Dalitz plot decomposition can be found in Ref.~\cite{Mikhasenko:2019rjf}. The final amplitude for a decay $A\to(B\to DE)C$ with defined helicities of all particles is
\begin{align}\label{eq:helAmpLS}
    \begin{split}
    &\qquad\mathcal{A} =
	D^{J_A}_{(\lambda_B-\lambda_C)\lambda_A}(\varphi_B,\theta_B,-\varphi_B)D^{J_B}_{(\lambda_D-\lambda_E)\lambda_B}(\varphi_D,\theta_D,-\varphi_D) \\
	&\times \sum_{L=|J_A-S|}^{|J_A+S|}\sum_{S=|J_B-J_\gamma|}^{|J_B+J_C|}\sum_{l=|J_B-s|}^{|J_B+s|}\sum_{s=|J_D-J_E|}^{|J_D+J_E|}\left[C^{A\to BC}_1C^{A\to BC}_2C^{B\to DE}_1C^{B\to DE}_2\right. \\
	&\qquad\qquad\qquad\qquad\qquad\qquad\qquad\qquad\left.\times H^{A\to BC}_{LS} (-1)^{J_C-\lambda_C} H^{B\to DE}_{ls} (-1)^{J_E-\lambda_E}\right. \\
	&\qquad\qquad\qquad\qquad\qquad\qquad\qquad\qquad\left. \times X(m_{DE})\right] \ ,
	\end{split}
\end{align}
where the couplings $H$ are now in the $LS$ basis. The Clebsch-Gordan coefficient $C_1$ is the coefficient for the spin-spin coupling between the children and $C_2$ is the coefficient for the coupling of total spin to orbital angular momentum. To improve readability, the upper case $L,S$ refer to orbital angular momentum $L$ and total spin $S$ in the decay $A\to BC$ and lower case $l,s$ refer to the corresponding quantities in the decay $B\to DE$.

\subsection{The full decay rate}
To obtain the full matrix element, we sum coherently over all internal parameters such as the helicity states of the resonance, $\lambda_B$, and the number of different resonances $B$, as they cannot be directly measured and appear in superpositions. To obtain the full differential decay rate from this, we take the incoherent sum over the helicities of initial and final state particles, where the helicity of the parent $A$ is averaged with respect to its polarisation. The full differential decay rate is
\begin{align}\label{eq:rateGeneral}
\begin{split}
    &\derivative{\Gamma}{(\varphi_B,\theta_B,\varphi_D,\theta_D,m_{DE})} \\
    &\qquad\qquad= \sum_{\lambda_A}\rho_{\lambda_A}\sum_{\lambda_C}\sum_{\lambda_D}\sum_{\lambda_E}\left|\sum_{B}\sum_{\lambda_B}\mathcal{A}_{\lambda_A,\lambda_B,\lambda_C,\lambda_D,\lambda_E}^{J_A,J_B}(\varphi_B,\theta_B,\varphi_D,\theta_D,m_{DE})\right|^2 \ .
\end{split}
\end{align}
The coefficients $\rho_{\lambda_A}$ are the diagonal elements of the polarisation density matrix and encode the polarisation of the parent particle $A$ along the quantisation axis, which we chose to be the $z$ axis of the helicity frame of $A$.\footnote{This choice of quantisation axis is made in the main body of the text for ease of explanation. The axis is actually arbitrary; a discussion of this is provided in Appendix~\ref{app:helAmp}.}

The decay $\LbpKGam$ is dominated by the decay via $\Lambda$ resonances, $\LSt$, decaying strongly into $\proton\Km$. However, the process $\Lb\to (P_c\to \proton\gamma)\Km$ via a charm pentaquark $P_c$, as in those observed by LHCb~\cite{Aaij:2015tga}, is also theoretically possible. The full decay rate using two decay chains $A\to(B\to DE)C$ and $A\to(F\to CD)E$ with helicity amplitudes $\mathcal{A}_B$ and $\mathcal{A}_F$, both defined according to Eq.~\eqref{eq:helAmpThreeBody} -- or Eq.~\eqref{eq:helAmpLS} equivalently -- is given by
\begin{equation*}
    \derivative{\Gamma}{(\Omega_B,\Omega_F,m_{DE},m_{CD})} = \sum_{\lambda_A}\rho_{\lambda_A}\sum_{\lambda_C}\sum_{\lambda_D}\sum_{\lambda_E}\left|\sum_{B}\sum_{\lambda_B}\mathcal{A}_B + \sum_{F}\sum_{\lambda_F}\mathcal{A}_F\right|^2 \ ,
\end{equation*}
where $\dd\Omega_B = \dd(\varphi_B,\theta_B,\varphi_D,\theta_D)$ represents the angular dependencies of the $B$ decay chain as defined in Section~\ref{sec:helFormThreeBody} and $\dd\Omega_F = \dd(\varphi_F,\theta_F,\varphi_C,\theta_C)$ corresponds to the angular dependencies of the $F$ decay chain, defined analogously. Nevertheless, there are still only two degrees of freedom in a three-body decay, such that those are not independent variables, but can be connected by an appropriate rotation.
For an example usage of two decay chains, consult the details of the discovery of pentaquark states in the decay $\Lb\to \proton\kaon^-\jpsi$ at LHCb~\cite{Aaij:2015tga}.
\section{The case of $\Lb\to(\LSt\to \proton\Km)\gamma$}
\label{sec:specificLbpKG}
In this section, we apply the general amplitude formalism introduced above to the decay $\Lb\to(\LSt\to \proton\Km)\gamma$. After first explaining how the amplitude can be simplified for this decay, the differential decay rate is given and subsequently examined.
\subsection{Simplifications}\label{sec:simplifications}
Given that the kaon particle has spin 0, the total spin $s$ in the decay $\LSt\to \proton\Km$ is equal to the proton spin $J_p=1/2$. According to the rules of angular momentum coupling, the orbital angular momentum $l$ and the total spin $s$ must satisfy the condition $|l-s|\leq J_{\LSt}\leq|l+s|$, which restricts the possible orbital angular momenta to two consecutive integers: $l=J_{\LSt}\pm1/2$. Additionally, the decay $\LSt\to \proton\Km$ is a strong process requiring parity conservation, explicitly
\begin{equation*}
    P_{\LSt} = P_pP_K(-1)^l \ ,
\end{equation*}
where $P_i$ is the parity of particle $i$.
The lack of spin-spin coupling between proton and kaon and parity conservation in the strong decay $\LSt\to \proton\Km$ fix the orbital angular momentum $l$ between the proton and the kaon unambiguously.

In summary, there is only one $ls$ coupling parameter, $H^{\LSt\to \proton\Km}_{ls}$, for each $\LSt$ resonance, which enters as an overall factor to this resonance contribution to the amplitude. Subsequently, a fit to the data cannot determine their value, and we choose to absorb them into the $LS$ couplings $H^{\Lb\to\LSt\gamma}_{LS}$.

\subsection{Differential decay rate}\label{sec:diffDecRate}
The helicity amplitude in Eq.~\eqref{eq:helAmpLS} after the simplifications reads
\begin{align}\label{eq:helAmpLb}
    \begin{split}
    \mathcal{A} &= D^{J_{\Lb}}_{(\lambda_{\LSt}-\lambda_\gamma)M_{\Lb}}(\varphi_{\LSt},\theta_{\LSt},-\varphi_{\LSt})D^{J_{\LSt}}_{\lambda_p\lambda_{\LSt}}(\varphi_p,\theta_p,-\varphi_p) \\
	&\qquad\times \sum_{L=|J_{\Lb}-S|}^{|J_{\Lb}+S|}\sum_{S=|J_{\LSt}-J_\gamma|}^{|J_{\LSt}+J_\gamma|}C^{\Lb\to\LSt\gamma}_1C^{\Lb\to\LSt\gamma}_2C^{\LSt\to \proton\Km}_2\times H^{\Lb\to\LSt\gamma}_{LS}\times X(m_{\proton\Km}) \ .
	\end{split}
\end{align}
Substituting this amplitude into the general decay rate of Eq.~\eqref{eq:rateGeneral} yields
\begin{align}\label{eq:rateLb}
\begin{split}
    &\derivative{\Gamma}{(\varphi_{\LSt},\theta_{\LSt},\varphi_p,\theta_p,m_{\proton\Km})} \\
    &\qquad\qquad= \sum_{\lambda_{\Lb}}\rho_{\lambda_{\Lb}}\sum_{\lambda_\gamma}\sum_{\lambda_p}\left|\sum_{\LSt}\sum_{\lambda_{\LSt}}\mathcal{A}_{\lambda_{\Lb},\lambda_{\LSt},\lambda_\gamma,\lambda_p,0}^{J_{\Lb},J_{\LSt}}(\varphi_{\LSt},\theta_{\LSt},\varphi_p,\theta_p,m_{\proton\Km})\right|^2 \ .
\end{split}
\end{align}
Note that there is no sum over the kaon states because $J_K=\lambda_K=0$. Since the spin of the $\Lb$ baryon is $J_{\Lb}=1/2$, its helicities can only take the values $\lambda_{\Lb}=\pm1/2$. This enables the coefficient $\rho_{\lambda_{\Lb}}$ to be parameterised in terms of the $\Lb$ polarisation:
\begin{equation*}
    \rho_{\pm1/2} = \frac{1\pm P_{\Lb}}{2} \ .
\end{equation*}

As a final comment in this section, we point out that the photon is massless and therefore cannot carry longitudinal polarisation, $\lambda_\gamma=0$, which reduces the number of possible helicity couplings. The basis transformation from helicity to $LS$ couplings however is a purely mathematical operation without prior physics knowledge, resulting in a set of $LS$ couplings that are not independent of each other. More explicitly, the formalism assumes that all helicity states $-\lambda_\gamma\leq J_\gamma\leq\lambda_\gamma$, i.e. $\lambda_\gamma=0,\pm1$, are present. An additional constraint like $\lambda_\gamma\not=0$ needs to be introduced manually. This is straightforward for the helicity couplings: $H_{\lambda_{\LSt}0}=0$. Due to the nontrivial relation between the helicity and $LS$ couplings, no individual $LS$ coupling can be set to zero, but rather the sums over $LS$ couplings that correspond to $H_{\lambda_{\LSt}0}$ can.

\subsection{Differential decay rate after evaluation of the helicity sums}
After executing all the helicity sums in Eq.~\eqref{eq:rateLb}, the differential decay rate reads
\begin{equation}\label{eq:decayRate}
	\derivative{\Gamma}{(\Omega,m_{\proton\Km})} = \sum_n \left(S_{1/2}^n+S_{3/2}^n+S_\text{mix}^n\right) 
	+ \sum_n\sum_{m<n}\left(T_{1/2}^{nm}+T_{3/2}^{nm}+T_\text{mix}^{nm}\right) \ ,
\end{equation}
divided into self-interaction terms $S$ and interference terms $T$, where the lower index describes which helicity states are interfering. For example, the term $S_{1/2}^n$ contains the interaction of the helicity $\pm1/2$ of a resonance with itself, whereas $T_{\text{mix}}^{nm}$ contains the interaction of the helicity state $\pm1/2$ of one resonance with the helicity state $\pm3/2$ of another resonance. This is a result of the conservation of the spin projection encoded into the indices of the Wigner-D functions. Explicitly, $|\lambda_{\Lb}|=|\lambda_{\LSt}-\lambda_\gamma|$ needs to be satisfied allowing only $\lambda_{\LSt}=\pm1/2,\pm3/2$ and requiring the helicities of $\LSt$ and the photon to have the same sign. The sums in Eq.~\eqref{eq:decayRate} are taken over the various $\LSt$, i.e. $m,n=\LSt_1,\LSt_2,...$

The explicit form of the interaction terms are given in Appendix~\ref{app:sums}. However, in the following, some features of the full differential decay rate are illuminated. The angles $\theta_{\LSt},\varphi_{\LSt},\varphi_p$ only appear in products with the longitudinal $\Lb$ polarisation $P_b$.\footnote{The quantisation axis is arbitrary, but in the main body of this text we chose it to be parallel to the $\Lb$ spin. Details on this choice can be found in Appendix~\ref{app:helAmp}.} Since the $\Lb$ is produced in a strong interaction process, the longitudinal polarisation must be zero. Despite cancelling the dependency on three of the decay angles, setting the longitudinal polarisation to zero in the decay rate does not decrease the degrees of freedom because the angles are kinematic variables. And in a three-body decay there are only two kinematic degrees of freedom, commonly represented by the standard Dalitz plane $(m_{p\gamma}^2,m_{\proton\Km}^2)$\cite{Dalitz:1953cp}; other choices such as the rectangular Dalitz plane $(\cos(\theta_p),m_{\proton\Km})$ are possible and in use.

In the decay $\Lb\to(\LSt\to \proton\Km)\jpsi$, the most abundant $\LSt$ contributions consist of various resonances with spin 1/2 and one resonance with spin 3/2 \cite{Aaij:2015tga}. Assuming this is also true for the decay with a photon, the shape of the differential decay rate in $\cos \theta_p$ is dominated by a second order polynomial; further information can be found in Appendix~\ref{app:contributions}.

\section{Fitting the decay rate to data}
\label{sec:fitstodata}
Here we describe several aspects related to the fit of the amplitude model to measured data. Explicitly, this section begins with a brief discussion of the treatment of detector effects, followed by possible reductions of the number of fit parameters for computational improvements, and concludes with presenting a possible choice of the observables determined by the fit.
\subsection{Detector effects}
However well the model describes the physics of the decay of interest, when looking at experimental data, detector effects need to be considered. Most generally, a measured distribution $h$ corresponds to the convolution of the true distribution, described by a model $f$, with the detector response, described by a function $g$:
\begin{equation*}
    h(y) = \int_{-\infty}^\infty f(y-x)g(x)\dd x \ .
\end{equation*}
Recovering the truth $f$ through unfolding is a highly nontrivial problem. Quite commonly, this task is approached by splitting the detector response into two categories: acceptance and resolution. The former accounts for asymmetries 
in the efficiency over the phase space introduced by the detector geometry and the selection requirements, 
while the latter describes the finite precision of a measurement, that might depend on the measured value itself. 
Both effects are typically studied with simulated events, on which the full detector response has been emulated,
and can be crosschecked using dedicated control modes with similar topology to the decay of interest.
A detailed example can be found in Ref.~\cite{LHCb-PAPER-2015-051}.

The modelling of the acceptance is particular to each experimental setup and can also vary between analyses of the same data depending on their aims. Nevertheless, a general comment on the treatment of resolution effects can be made. As mentioned in the previous section, a three-body decay such as $\LbpKGam$ has only two kinematic degrees of freedom, such that the fit is also performed in two dimensions. Commonly, the invariant-mass-squared of two combinations of the child particles, for example $m_{\proton\Km}^2$ and $m_{p\gamma}^2$, are chosen to represent those degrees of freedom. In this plane -- in the previous section denoted as the standard Dalitz plane -- the phase space has highly nonlinear boundaries. Subsequently, the estimation of the resolution close to the edges of the phase space requires extra care. In such a case, moving to a parametrisation of the phase space that has less complex or even rectangular borders could be an advantage. An example of a rectangular Dalitz plane is the combination of $m_{\proton\Km}$ and $\cos(\theta_p)$. The transformation is given by
\begin{equation*}
    m_{p\gamma}^2 = m_p^2 + 2\frac{m_{\Lb}}{m_{\proton\Km}}\left(E_p^{\proton\Km}E_\gamma^{\Lb}+\left|\vec{p}_p^{~\proton\Km}\right|\left|\vec{p}_\gamma^{~\Lb}\right|\cos(\theta_p)\right) \ ,
\end{equation*}
where the lower index of the energies $E$ and momenta $\vec{p}$ refer to the particle and the upper index refers to the rest frame in which they are measured. Additional comments on the transformation are found in Appendix~\ref{app:transformation}.
\subsection{Reduction of number of fit parameters}
The decay rate in Eq.~\eqref{eq:decayRate} has four (six) complex fit parameters $H_{LS}^{\Lb\to\LSt\gamma}$ for each $\LSt$ with spin $J_{\LSt}=1/2$ ($J_{\LSt}\geq3/2$). This may result in a computationally expensive and potentially unstable fit because the $\proton\Km$ spectrum is very rich in resonances.

If necessary, a reduction of parameters can be achieved by neglecting high orbital angular momentum 
contributions. The fractions of momentum over mass in the orbital angular momentum barrier factors,
\begin{equation*}
    \left(\frac{q}{m_{\Lambda_{\text{b}}}}\right)^L\left(\frac{p}{m_{\Lambda^*}}\right)^l ,
\end{equation*}
are always smaller than one, such that the contributions for high orbital angular momentum $L$ or $l$ may be negligible, assuming that the $LS$ couplings for one resonance are of similar magnitude.

\begin{figure}
    \centering
    \includegraphics[width=.8\textwidth]{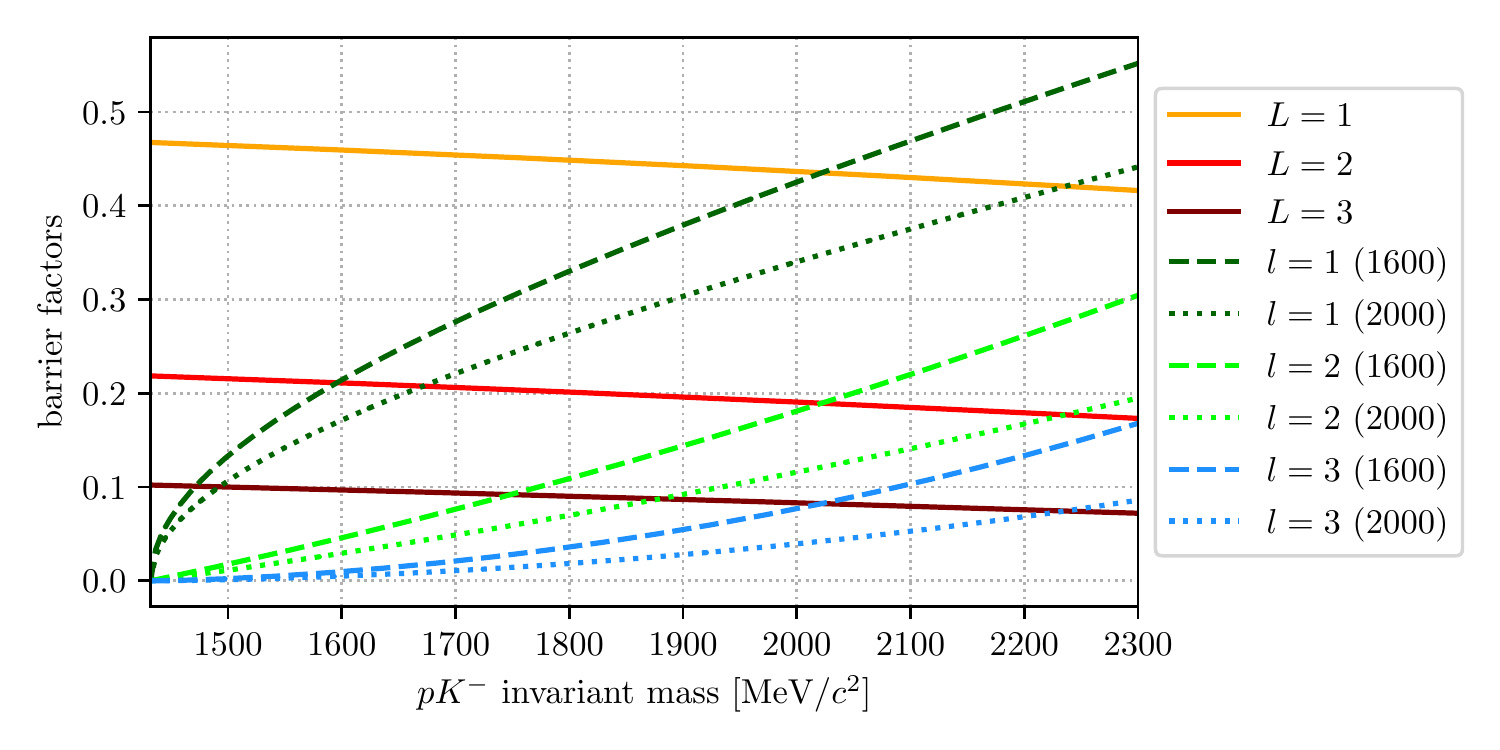}
    \caption{Barrier functions for different resonance masses. Following the notation of before, capital $L$ corresponds to the orbital angular momentum in the $\Lb$ decay and minuscule $l$ belongs to the $\LSt$ decay. As the barrier functions in the $\LSt$ decay depend on the mass of the decaying resonance, for each orbital angular momentum $L$, two example masses are plotted: $1600\mevcc$ and $2000\mevcc$.}
    \label{fig:barriers}
\end{figure}

In the particular case of $\Lb\to(\LSt\to \proton\Km)\gamma$, the orbital angular momentum $l$ between the proton and the kaon is fixed for each $\LSt$, as outlined in Section~\ref{sec:simplifications}. However, various $L$ are possible for each resonance. As the photon is massless, its momentum $q$ may be very high, such that the fraction $q/m_{\Lb}$ -- while still always being smaller than one -- does not decrease much with higher $L$. Fig.~\ref{fig:barriers} displays the magnitude of the barrier factors for different $L,l$ and resonance mass for this case. For a fixed $l$, the suppression is largest for heavy resonances with large spin (resonances with large spin decay to a proton kaon pair with large $l$). The angular momentum $L$ can take different values, resulting in different suppression factors. As an example, the contribution for $L=3$ is suppressed by a factor of {\footnotesize$\lesssim$}$ 10 \% $. This suppression would be significantly larger if the photon is replaced by a massive particle.

In addition, and as mentioned previously in Section~\ref{sec:diffDecRate}, the coupling parameters $H_{LS}^{\Lb\to\LSt\gamma}$ are not independent due to the lack of longitudinal photon polarization $\lambda_\gamma=0$. Thus by including those dependencies in the fit, the number of free parameters is reduced by two for each $\LSt$, leaving only two (four) parameters for each $\LSt$ with spin $1/2$ ($\geq3/2$).
\subsection{Fit fractions}
The primary aim of the amplitude analysis of the decay $\Lb\to(\LSt\to \proton\Km)\gamma$ is to understand the $\LSt$ resonance spectrum. The fit fractions
\begin{equation*}
    f_n = \frac{\int_{\mathcal{D}}\sum_{\lambda_{\Lb}\lambda_{\g}\lambda_p}\left|\mathcal{M}_n\right|^2\dd\mathcal{D}}{\int_{\mathcal{D}}\sum_{\lambda_{\Lb}\lambda_{\g}\lambda_p}\left|\sum_n\mathcal{M}_n\right|^2\dd\mathcal{D}}
\end{equation*}
can serve as model-independent observables. Here, the amplitude of one resonance, $\mathcal{M}_n$, is given by the sum over all its helicitiy amplitudes described in Eq.~\eqref{eq:helAmpLb}: $\mathcal{M}_n=\sum_{\lambda_{\LSt_n}}\mathcal{A}$. The integral is evaluated over the Dalitz plane $\mathcal{D}$. Due to constructive and destructive interferences, the fit fractions $f_n$ do not necessarily add up to one. The interference fit fractions can be quantified using
\begin{equation*}
    i_{nm} = \frac{\int_{\mathcal{D}}\sum_{\lambda_{\Lb}\lambda_{\g}\lambda_p}\text{Re}\left(\mathcal{M}_n\mathcal{M}_m^*\right)\dd\mathcal{D}}{\int_{\mathcal{D}}\sum_{\lambda_{\Lb}\lambda_{\g}\lambda_p}\left|\sum_n\mathcal{M}_n\right|^2\dd\mathcal{D}} \ .
\end{equation*}
The interference fit fraction should be zero for resonances with different quantum numbers, a property that can be used to crosscheck the validity of the fit.
The fit fractions and interference fit fractions together always sum to one by definition.
A discussion of resonance interferences can be found in Ref.~\cite{Azimov:2009ta}.
\subsection{Additional measurements}
The decay \LbpKGam is dominated by the $\Lb\to\LSt\g$, $\LSt\to\proton\Km$ decay chain discussed in Section~\ref{sec:specificLbpKG}. Nevertheless, other resonances such as \Sigmaz resonances, $\Sigmaz^*$, can be present in the $\proton\Km$ spectrum. The decay can also proceed via other decay chains such as \mbox{$\Lb\to (P_c\to \proton\gamma)\Km$}. Generally, adding such contributions into the fit model allows the respective amplitudes to be measured.

In this specific case, $\Sigmaz^*$ contributions are expected to be minor since 
they proceed through a $\Delta I=1$ transition, contrary to the $\Delta I=0$ decays $\Lb\to\LSt\g$~\cite{Dery:2020lbc}. Although isospin suppressed, $\Sigmaz^*$ resonances could be present in the decay. If the amount of data is large enough to resolve these small contributions, it becomes necessary to include them to have the best possible description of the observed spectrum. 

The observation of intermediate $P_c$ states decaying to the $\proton\g$ 
final state would provide a confirmation of their resonant nature~\cite{Wang:2015jsa}
and a measurement of their currently unknown coupling to photons.

\section{Conclusions}
\label{sec:conclusions}
Following the recent measurement of LU using \LbpKll decays by LHCb, 
the study of the \proton\Km spectrum in this transition is required in order to interpret the result in terms of New Physics models. 
The radiative mode \LbpKGam, less suppressed and thus more abundant in data, provides an ideal benchmark to measure 
the \proton\Km spectrum at $q^2 = 0$.

Summarising existing literature, a derivation of the differential decay rate for a general three body decay is presented assuming the decay proceeds via two two-body decays. This is followed by its application to the decay $\LbpKGam$. Finally, particularities of a fit to data including possible choices of variables and the reduction of fit parameters are discussed.

This information can be used to perform an amplitude analysis of the \LbpKGam transition for the first time, exploiting the LHCb dataset.

\section*{Acknowledgements}
%
%
\noindent The authors would like to express their gratitude to  Anton Poluektov, Mikhail Mikhasenko and Liming Zhang for useful discussions. Special thanks go to Anton Poluektov and Donal Hill for their feedback on the manuscript. 
J. Albrecht gratefully acknowledges support of the European Research Council, ERC Starting Grant: PRECISION 714536. C. Marin Benito is supported by the ANR grant BACH. A. Beck is supported by the Erasmus+ program of the European Commission.

\clearpage

\appendix
\section{Formal derivation of the helicity amplitude}\label{app:helAmp}
The derivation in this appendix chapter closely follows the explanations of a comprehensive guide on the helicity formalism \cite{Richman:1984gh}. \\
In a decay $A\to BC$, the state of the parent particle in its rest frame is fully described by its quantum numbers $J_A$ and $M_A$, where $J_A$ is the parent's spin and $M_A$ is its spin projection onto the $z$-axis. Note that the choice of this initial coordinate system and thus also the quantisation axis $z$ is arbitrary but must be fixed throughout the computations. The combined state of the child particles in the initial coordinate system is completely defined by their momenta $\vec{p}_B,\vec{p}_C$ and helicities $\lambda_B,\lambda_C$. The amplitude of the decay is
\begin{equation*}
	\mathcal{A}(\vec{p}_B,\vec{p}_C,\lambda_B,\lambda_C,J_A,M_A) = \bra{\vec{p}_B,\lambda_B;\vec{p}_C,\lambda_C}U\ket{J_AM_A}
\end{equation*}
where $U$ is the time evolution operator. Instead of using the conventional Cartesian momenta of $B$ and $C$, one may as well use the polar and azimuthal angle of the decay axis $\theta_B,\varphi_B$ and the magnitude $p$ of the momentum in the parent rest frame:
\begin{equation}\label{eq:amp}
	\mathcal{A}(\vec{p}_B,\vec{p}_C,\lambda_B,\lambda_C,J_A,M_A) = \bra{p,\theta_p,\varphi_p,\lambda_B,\lambda_C}U\ket{J_AM_A} \ .
\end{equation}
The magnitude of the momentum can be factored out, as it is fixed by four momentum conservation
\begin{align}\label{eq:momentum}
	&\left(p_A\right)^\mu\left(p_A\right)_\mu = \left(p_B+p_C\right)^\mu\left(p_B+p_C\right)_\mu\notag \\
	\Leftrightarrow\quad &p = \sqrt{\frac{(M_A^2-(M_B+M_C)^2)(M_A^2-(M_B-M_C)^2)}{4M_A^2}}
\end{align}

To make use of the helicity formalism, we need to rewrite the child state in terms of the child spins $J_B,J_C$. To this end, a final (in contrast to initial) coordinate system $F$ is defined, where the $z$-axis is parallel to the momentum of one of the children $\vec{p}_B$. The transformed state is
\begin{align}\label{eq:transform}
	\ket{\theta_B,\varphi_B,\lambda_B,\lambda_C}_I \rightarrow \ket{J_Bm_B;J_Cm_C}_F \ ,
\end{align}
where instead of the helicities $\lambda_B,\lambda_C$, for the state in the final coordinate system $F$, the projections of the spins onto $z_F$, i.e. $m_B=\lambda_B,m_C=-\lambda_C$, are used.

We rotate the parent particle from $I$ to $F$
\begin{align}\label{eq:rotparent}
	R(\alpha,\beta,\gamma)\ket{J_AM_A}_I = \sum_{M'=-J_A}^{J_A} D_{M'M_A}^{J_A}(\alpha,\beta,\gamma)\ket{J_AM'}_F \ .
\end{align}
Appendix~\ref{app:rotations} explains how to carry out this rotation. To actually end up in $F$, the rotation angles must be $\alpha=\varphi_B$ and $\beta=\theta_B$. After these two rotations, the direction that was $z$ in the initial frame, now coincides with the direction of $\vec{p}_B$ which is the $z$-axis in $F$. The third rotation of $R(\alpha,\beta,\gamma)$ rotates around $z$ in $F$. The states in $F$ though are eigenstates under such transformations by definition. Therefore the third angle $\gamma$ has no physical meaning and is often set to $\gamma=0$ or $\gamma=-\varphi_B$.

By inserting the transformations given by Eqs.~\eqref{eq:transform} and \eqref{eq:rotparent}, the amplitude in Eq.~\eqref{eq:amp} can now be evaluated in the final coordinate system and reads
\begin{equation*}
    \mathcal{A}(p_B,\varphi_B,\theta_B,\lambda_B,\lambda_C,J_A,M_A) = \sum_{M'=-J_A}^{J_A} D_{M'M_A}^{J_A}(\varphi_B,\theta_B,-\varphi_B)\bra{p}\bra{J_Bm_B;J_Cm_C}U\ket{J_AM'} \ .
\end{equation*}
The spin states are scalars under the time evolution operator $U$. Introducing $\ket{p_A=0}$ to describe the dynamics of the parent particle $A$ in its rest frame, the amplitude becomes
\begin{align*}
    \mathcal{A}(p_B,\varphi_B,\theta_B,\lambda_B,\lambda_C,J_A,M_A) &= \sum_{M'=-J_A}^{J_A} D_{M'M_A}^{J_A}(\varphi_B,\theta_B,-\varphi_B)\braket{J_Bm_B;J_Cm_C}{J_AM'}\bra{p}U\ket{0} \notag\\
    &= D_{(\lambda_B-\lambda_C)M_A}^{J_A}(\varphi_B,\theta_B,-\varphi_B)H_{\lambda_B\lambda_C} \ .
\end{align*}
Here we have introduced the helicity coupling
\begin{equation}\label{eq:helAmpApp}
    H_{\lambda_B\lambda_C}=\braket{J_B\lambda_B;J_C-\lambda_C}{J_A(\lambda_B-\lambda_C)}\bra{p}U\ket{0} \ .
\end{equation}

Choosing the initial coordinate system defines the quantisation axis $z$. Commonly, the intital coordinate system is the helicity frame of $A$. The spin projection $M_A$ of the parent then equals the helicity of the parent, because $z$ coincides with the direction of the parent momentum. This is used in the main text and shown in Fig.~\ref{fig:rotationHelFrame}. In this case, the longitudinal polarisation is measured.

Another useful option can be to choose $z$ perpendicular to the momentum of the parent. In this case, the spin projection $M_A$ is different to the helicity of the parent. This means that instead of summing all parent helicities in Eq.~\eqref{eq:rateGeneral}, we sum all possible values of $M_A$ and the polarisation $P_A$ (encoded in the density matrix $\rho_A$) becomes the transverse polarisation. Choosing this coordinate system does not change the structure of the amplitude. However, additionally to changing the interpretation of $M_A$ and $P_A$, the computation of the decay angles in the decay $A\to BC$, $\varphi_B$ and $\theta_B$, differs. For example, the polar angle $\theta_B$ in the longitudinal case is the angle between $A$ and $B$ momentum directions. In the transverse case it is the angle between the polarisation axis and the $B$ momentum. The angles in a consecutive decay do not change.
\section{Dynamics}\label{app:dynamics}

As described in the main text, the common parametrisation of the  dynamical part $X(m_{DE})$ depends on the orbital angular momenta between the children in each decay. This is why we transform the helicity couplings to $LS$ couplings. This consists of two main steps: first, coupling the spins of the two children in each of the decays to a total spin $S$ and second, couple the total spin to the orbital angular momentum $L$ between the two children in each decay.

The coupling of the children spins $J_B,J_C$ in a decay $A\to BC$ to a total spin $S$ is
\begin{align}\label{eq:spinspin}
	\ket{J_B\lambda_B;J_C-\lambda_C} &= \sum_{s=|J_B-J_C|}^{|J_B+J_C|}\sum_{m_S=-S}^S\braket{Sm_S,J_BJ_C}{J_B\lambda_B;J_C-\lambda_C}\ket{Sm_S,J_BJ_C}_F\notag \\
	&= \sum_{S=|J_B-J_C|}^{|J_B+J_C|}\underbrace{\braket{S(\lambda_B-\lambda_C),J_BJ_C}{J_B\lambda_B;J_C-\lambda_C}}_{C_1}\ket{S(\lambda_B-\lambda_C),J_BJ_C} \ ,
\end{align}
where $C_1$ is the related Clebsch-Gordan coefficient.
To obtain the total angular momentum $J$, the total spin $S$ must be coupled to the orbital angular momentum $L$ between the children:
\begin{align}\label{eq:lsCoupling}
	\ket{Lm_L}\ket{S(\lambda_B-\lambda_C),J_BJ_C} &:= \ket{Lm_L;S(\lambda_B-\lambda_C)} \notag\\
	&= \sum_{J=|L-S|}^{|L+S|}\sum_{M_J=-J}^{J}\bra{JM_J,LS}\ket{Lm_L;S(\lambda_B-\lambda_C)}\ket{JM_J,LS} \notag\\
	&= \sum_{J=|L-S|}^{|L+S|}\underbrace{\braket{J(\lambda_B-\lambda_C),LS}{Lm_L;S(\lambda_B-\lambda_C)}}_{C_2}\ket{J(\lambda_B-\lambda_C),LS} \ .
\end{align}
Following the notation of before, $M_J$ is the projection of the total angular momentum $J$ onto the $z$ axis, $m_L$ is the projection of the orbital angular momentum $L$ and the related Clebsch-Gordan coefficient is called $C_2$. The projection of $L$ onto $z$ is zero, since the orbital angular momentum vector is always perpendicular to the momentum vector (which is parallel to $z$).

Inserting Eq.~\eqref{eq:lsCoupling} into Eq.~\eqref{eq:spinspin} gives a parametrisation of the child state in terms of orbital angular momentum $L$ and total spin $S$:
\begin{equation}\label{eq:totAngMom}
	\ket{J_B\lambda_B;J_C-\lambda_C} = \sum_L\sum_{S=|J_B-J_C|}^{|J_B+J_C|} \sum_{J=|L-S|}^{|L+S|}C_1C_2\ket{J(\lambda_B-\lambda_C),LS} \ .
\end{equation}
The helicity coupling in Eq.~\eqref{eq:helAmpApp} can now be rewritten to
\begin{align*}
    H_{\lambda_B\lambda_C} &= \braket{J_B\lambda_B;J_C-\lambda_C}{J_A(\lambda_B-\lambda_C)}\bra{p_B(m_{DE})}U\ket{0} \\
    &= \sum_L\sum_{S=|J_B-J_C|}^{|J_B+J_C|} \sum_{J=|L-S|}^{|L+S|}C_1C_2\braket{J(\lambda_B-\lambda_C),LS}{J_A(\lambda_B-\lambda_C)}\bra{p_B(m_{DE})}U\ket{0} \\
    &= \sum_L\sum_{S=|J_B-J_C|}^{|J_B+J_C|} \sum_{J=|L-S|}^{|L+S|}C_1C_2\delta_{JJ_A}\bra{p_B(m_{DE})}U\ket{0} \\
    &= \sum_{S=|J_B-J_C|}^{|J_B+J_C|} \sum_{L=|J_A-S|}^{|J_A+S|}C_1C_2\bra{p_B(m_{DE})}U\ket{0} \ .
\end{align*}
The product of both helicity couplings in two consecutive two-body decays including a dynamical function $X(m_{DE})$ is
\begin{align*}
    &\qquad H_{\lambda_B\lambda_C}H_{\lambda_D\lambda_E}X_{J_B}(m_{DE}) \\
    &= \sum_{S=|J_B-J_C|}^{|J_B+J_C|} \sum_{L=|J_A-S|}^{|J_A+S|}\sum_{s=|J_D-J_E|}^{|J_D+J_E|} \sum_{l=|J_B-s|}^{|J_B+s|}C_1^AC_2^AC_1^BC_2^B\bra{p_B(m_{DE})}U\ket{0}\bra{p_D(m_{DE})}U\ket{0} \\
    &= \sum_{S=|J_B-J_C|}^{|J_B+J_C|} \sum_{L=|J_A-S|}^{|J_A+S|}\sum_{s=|J_D-J_E|}^{|J_D+J_E|} \sum_{l=|J_B-s|}^{|J_B+s|}C_1^AC_2^AC_1^BC_2^BH_{LS}H_{ls}X_{Ll}(m_{DE}) \ .
\end{align*}
The dynamical function $X_{J_B}(m_{DE})$ is an unknown parametrisation in terms of spins and helicities while the dynamical function $X_{Ll}(m_{DE})$ is defined in the main text in Eq.~\eqref{eq:dynamics} depending on the angular momenta $L,l$.
\section{Rotations and Wigner-D functions}\label{app:rotations}
The line of argument in this appendix chapter is largely taken from a common reference on the helicity formalism~\cite{Richman:1984gh}. \\
The generators of rotations are angular momenta. An arbitrary rotation in 3D space can therefore be described by
\begin{equation}\label{eq:rotMat}
	R(\alpha,\beta,\gamma) = e^{-i\gamma \vec{J}\hat{r}_3}e^{-i\beta \vec{J}\hat{r}_2}e^{-i\alpha \vec{J}\hat{r}_1} \ ,
\end{equation}
where $\alpha,\beta,\gamma$ are the Euler angles, $\hat{r}_i$ are the unit vectors along the Euler rotation axes, and $\vec{J}$ is the angular momentum generating the rotation.

The aim is rotating a spin state $\ket{JM}$ where $M$ is the projection of $J$ onto the $z$ axis in the initial coordinate system, into a final coordinate system, where the projection of the spin onto the new quantisation axis $z'$ is called $M'$. The first step in performing the rotation is inserting a complete set of basis functions formed by the states $\ket{JM'}$ where $-J\leq M'\leq J$:
\begin{align*}
	R(\alpha,\beta,\gamma)\ket{JM} &= \sum_{M'=-J}^J\ket{JM'}\bra{JM'}R(\alpha,\beta,\gamma)\ket{JM} \\
	&\overset{\eqref{eq:rotMat}}{=} \sum_{M'=-J}^J\bra{JM'}e^{-i\gamma \vec{J}\hat{r^3}}e^{-i\beta \vec{J}\hat{r^2}}e^{-i\alpha \vec{J}\hat{r^1}}\ket{JM}\ket{JM'} \ .
\end{align*}
In the Euler rotation formalism, the first rotation axis $\hat{r}_1$ coincides with the $z$ axis of the initial system and the last rotation axis $\hat{r}_3$ coincides with the $z$ axis of the final system. Exploiting that $\ket{JM}$ and $\ket{JM'}$ are eigenstates under rotations around their respective $z$ axis, the rotation becomes
\begin{align*}
    R(\alpha\beta\gamma)\ket{JM} &= \sum_{M'=-J}^J\underbrace{\bra{JM'}e^{-i\gamma M'}e^{-i\beta \vec{J}\hat{r^2}}e^{-i\alpha M}\ket{JM}}_{D_{M'M}^J(\alpha,\beta,\gamma)}\ket{JM'} \\
	&= \sum_{M'=-J}^Je^{-i\gamma M'}\underbrace{\bra{JM'}e^{-i\beta \vec{J}\hat{r^2}}\ket{JM}}_{d_{M'M}^J(\beta)}e^{-i\alpha M}\ket{JM'} \ .
\end{align*}
Here, the commonly used Wigner-d and Wigner-D matrix elements have been identified.

\section{Transformation between Dalitz variables}\label{app:transformation}
In the following, the upper index of a variable refers to the rest frame in which the relevant quantity is evaluated and the lower index refers to the particle that the quantity is related to.

If $p_\gamma^{\Lb}$ is the photon four momentum in the $\Lb$ rest frame, a boost into the $\proton\Km$ rest frame transforms it to
\begin{equation*}
	p_\gamma^{\proton\Km} = p_\gamma^{\Lb}\frac{m_{\Lb}}{m_{\proton\Km}}
\end{equation*}
The invariant mass of the $p\gamma$ system evaluated in the $\proton\Km$ rest frame is
\begin{align*}
	m_{p\gamma}^2 &= \left(p_{\proton}^{\proton\Km}+p_\gamma^{\proton\Km}\right)_\mu\left(p_{\proton}^{\proton\Km}+p_\gamma^{\proton\Km}\right)^\mu \\
	&= m_{\proton}^2 + 2\left(p_{\proton}^{\proton\Km}\right)_\mu\left(p_\gamma^{\proton\Km}\right)^\mu \\
	&= m_{\proton}^2 + 2\left(E_{\proton}^{\proton\Km}E_\gamma^{\proton\Km}-\left|\vec{p}_{\proton}^{~\proton\Km}\right|\left|\vec{p}_\gamma^{~\proton\Km}\right|\cos\theta_{p\gamma}\right)
\end{align*}
In the $\proton\Km$ rest frame, the angle between proton and photon momentum is \mbox{$\cos\theta_{p\gamma} = \cos(180-\theta_{\proton})=-\cos\theta_{\proton}$}, where $\theta_{\proton}$ is the angle between the proton and $\LSt$ momentum directions, ie. $\theta_{\proton}$ is the helicity angle of the proton.

Replacing the angle and inserting the boost of the photon momentum gives the transformation between $m_{p\gamma}$ and $\cos\theta_{\proton}$ in terms of measurable quantities:
\begin{align*}
	m_{p\gamma}^2 &= m_{\proton}^2 + 2\frac{m_{\Lb}}{m_{\proton\Km}}\left(E_{\proton}^{\proton\Km}E_\gamma^{\Lb}+\left|\vec{p}_{\proton}^{~\proton\Km}\right|\left|\vec{p}_\gamma^{~\Lb}\right|\cos\theta_{\proton}\right) \ .
\end{align*}
Assuming a decay only proceeds via phase space availability, the two dimensional distribution of the standard Dalitz variables $(m^2_{\proton\Km},m^2_{p\gamma})$ is uniform. To preserve the flatness of the phase space distribution when transforming into the rectangular Dalitz plane $(m_{\proton\Km},\cos\theta_{\proton})$, each event needs to be given the weight $1/\left(\left|\vec{p}_{\proton}^{~\proton\Km}\right|\left|\vec{p}_\gamma^{~\Lb}\right|\right)$ stemming from the Jacobian of the transformation:
\begin{align*}
    \dd m_{\proton\Km}^2\dd m_{p\gamma}^2
    &\propto \left|\vec{p}_{\proton}^{~\proton\Km}\right|\left|\vec{p}_\gamma^{~\Lb}\right|\dd m_{\proton\Km}\dd\cos\theta_{\proton} \ .
\end{align*}
\section{Interaction terms}\label{app:sums}
Notation: $J_n$ is the spin of $n$ and $J_m$ is the spin of $ m$, similarly $P_{n,m}$ is the parity of $n$ or $m$.
\begin{align*}
	S_{1/2}^n &= \frac{1}{2}\left[\left(1-P_b\cos\theta_{\LSt}\right)\mathcal{Q}_{1/21}^{nn} + \left(1+P_b\cos\theta_{\LSt}\right)\mathcal{Q}_{-1/2-1}^{nn}\right] \\
	&\quad\times\left[\left(d^{J_n}_{1/21/2}(\theta_p)\right)^2 + \left(d^{J_n}_{1/2-1/2}(\theta_p)\right)^2\right]
	\\
	S_{3/2}^n &= \frac{1}{2}\left[\left(1+P_b\cos\theta_{\LSt}\right)\mathcal{Q}_{3/21}^{nn} + \left(1-P_b\cos\theta_{\LSt}\right)\mathcal{Q}_{-3/2-1}^{nn}\right] \\
	&\quad\times\left[\left(d^{J_n}_{3/21/2}(\theta_p)\right)^2 + \left(d^{J_n}_{3/2-1/2}(\theta_p)\right)^2\right] \\
	S_\text{mix}^n &= P_b\sin(\theta_{\LSt})\text{Re}\left[
	\exp\left(i(\varphi_{\LSt}-\varphi_p)\right)\left(-\mathcal{Q}_{1/23/21}^{nn} + \mathcal{Q}_{-3/2-1/2-1}^{nn}\right)
	\right] \\
	&\quad\times\left[d^{J_n}_{1/21/2}(\theta_p)
	d^{J_n}_{3/21/2}(\theta_p) + d^{J_n}_{1/2-1/2}(\theta_p)
	d^{J_n}_{3/2-1/2}(\theta_p)\right] \\
	T_{1/2}^{nm} &= \left[d^{J_n}_{1/21/2}(\theta_p)
	d^{J_m}_{1/21/2}(\theta_p) + P_nP_m(-1)^{J_n+J_m+1}d^{J_n}_{1/2-1/2}(\theta_p)
	d^{J_m}_{1/2-1/2}(\theta_p)\right] \\
    &\quad\times \left[\left(1-P_b\cos\theta_{\LSt}\right)\text{Re}\left(\mathcal{Q}_{1/21}^{nm}\right) + P_nP_m(-1)^{J_n+J_m+1}\left(1+P_b\cos\theta_{\LSt}\right)\text{Re}\left(\mathcal{Q}_{-1/2-1}^{nm}\right)\right] \\
	T_{3/2}^{nm} &= \left[d^{J_n}_{3/21/2}(\theta_p)
	d^{J_m}_{3/21/2}(\theta_p) + P_nP_m(-1)^{J_n+J_m+1}d^{J_n}_{3/2-1/2}(\theta_p)
	d^{J_m}_{3/2-1/2}(\theta_p)\right] \\
	&\quad\times\left[\left(1-P_b\cos\theta_{\LSt}\right)\text{Re}\left(\mathcal{Q}_{3/21}^{nm}\right) + P_nP_m(-1)^{J_n+J_m+1}\left(1+P_b\cos\theta_{\LSt}\right)\text{Re}\left(\mathcal{Q}_{-3/2-1}^{nm}\right)\right] \\
	T_\text{mix}^{nm} &= P_b\sin(\theta_{\LSt})\left[d^{J_n}_{1/21/2}(\theta_p)
	d^{J_m}_{3/21/2}(\theta_p) + P_nP_m(-1)^{J_n+J_m+1}d^{J_n}_{1/2-1/2}(\theta_p)
	d^{J_m}_{3/2-1/2}(\theta_p)\right] \\
	&\quad\times\left[-\text{Re}\left(\exp\left(i(\varphi_{\LSt}-\varphi_p)\right)\mathcal{Q}_{1/23/21}^{nm}\right)\right. \\
	&\qquad\qquad\qquad\left.+ P_nP_m(-1)^{J_n+J_m+1}\text{Re}\left(\exp\left(-i(\varphi_{\LSt}-\varphi_p)\right)\mathcal{Q}_{-1/2-3/2-1}^{nm}\right)\right] \\
	& +P_b\sin(\theta_{\LSt}) \left[d^{J_n}_{3/21/2}(\theta_p)
	d^{J_m}_{1/21/2}(\theta_p) + P_nP_m(-1)^{J_n+J_m+1}d^{J_n}_{3/2-1/2}(\theta_p)
	d^{J_m}_{1/2-1/2}(\theta_p)\right] \\
	&\quad\times\left[-\text{Re}\left(\exp\left(-i(\varphi_{\LSt}-\varphi_p)\right)\mathcal{Q}_{3/21/21}^{nm}\right)\right. \\
	&\qquad\qquad\qquad\left.+ P_nP_m(-1)^{J_n+J_m+1}\text{Re}\left(\exp\left(i(\varphi_{\LSt}-\varphi_p)\right)\mathcal{Q}_{-3/2-1/2-1}^{nm}\right)\right]
\end{align*}
The dynamical factors $\mathcal{Q}$ are the products of the helicity parameters that were transformed into the $LS$ basis:
\begin{align*}
	\mathcal{Q}^{nm}_{\lambda_n\lambda_m} &= \sum_{LS}C^{\Lb\to n\gamma}_1C^{\Lb\to n\gamma}_2C_2^{n\to \proton\Km} H^{\Lb\to n\gamma}_{L_nS_n}X_{L_nS_n}(m_{\proton\Km}) \\
	&\times \left(\sum_{L_mS_m}C^{\Lb\to m\gamma}_1C^{\Lb\to m\gamma}_2C_2^{m\to \proton\Km} H^{\Lb\to m\gamma}_{L_mS_m}X_{L_mS_m}(m_{\proton\Km})\right)^* \\
	&= \sum_{L_nS_nL_mS_m}C^{\Lb\to n\gamma}_1C^{\Lb\to n\gamma}_2C_2^{n\to \proton\Km}C^{\Lb\to m\gamma}_1C^{\Lb\to m\gamma}_2C_2^{m\to \proton\Km} \\
	&\qquad\qquad\qquad\times H^{\Lb\to n\gamma}_{L_nS_n}X_{L_nS_n}(m_{\proton\Km})\left( H^{\Lb\to m\gamma}_{L_mS_m}X_{L_mS_m}(m_{\proton\Km})\right)^*
\end{align*}

\section{Most dominant contributions in $\theta_p$}\label{app:contributions}
The self-interaction terms of the spin-1/2 resonances are flat in $\cos\theta_p$: $S_{1/2}^n(\theta_p)\propto 1$. The self-interaction term of the spin-3/2 resonance contributes a parabola centred around $\cos\theta_p=0$:
\begin{equation*}
    S_{1/2}^n(\theta_p)+S_{3/2}^n(\theta_p)+S_{\text{mix}}^n(\theta_p)\propto1+k\cos^2\theta_p \ .
\end{equation*}
The interferences between the spin-1/2 resonances are
\begin{equation*}
    T_{1/2}^{nm} \propto
    \begin{cases}
        1 & \text{same parity} \\
        \cos\theta_p & \text{opposite parity}
    \end{cases} \ .
\end{equation*}
And finally, the interferences between the spin-1/2 and the spin-3/2 resonance are
\begin{equation*}
    T_{1/2}^{nm} \propto
    \begin{cases}
        1+k\cos^2\theta_p & \text{same parity} \\
        \cos\theta_p & \text{opposite parity}
    \end{cases} \ .
\end{equation*}
Summarizing, the differential decay rate $\derivative{\Gamma}{(\cos\theta_p)}$ is expected to have a parabolic shape.

\addcontentsline{toc}{section}{References}
\bibliographystyle{LHCb}
\bibliography{main,standard,LHCb-PAPER,LHCb-CONF,LHCb-DP,LHCb-TDR}

\end{document}